\documentclass[fleqn,twoside]{article}
\usepackage{espcrc2}

\usepackage{graphicx}
\usepackage[figuresright]{rotating}

\newcommand{\AmS}{{\protect\the\textfont2
  A\kern-.1667em\lower.5ex\hbox{M}\kern-.125emS}}

\title{ $\nu_\mu$ disappearance at the SPL, T2K-I and the Neutrino Factory}

\author{A. Donini\address[UAM]{Departamento F\'{\i}sica Te\'orica e IFT, 
                               Universidad Autonoma Madrid, 28049 Madrid, Spain}
        \thanks{The authors acknowledge the financial support of MEC through project FPA2003-04597 
                and of the European Union through the networking activity BENE.},
        E. Fern\'andez-Mart\'{\i}nez\addressmark[UAM]
        and 
        S. Rigolin\addressmark[UAM]}
       
\begin{document}

\begin{abstract}
We study the $\nu_\mu$ disappearance channel at T2K-phase I and the SPL and analyse the achievable reduction 
of present uncertainties in $\theta_{23}$ and $\Delta m^2_{23}$. 
We analyse the impact of discrete ambiguities in sign($\Delta m^2{23}$) and sign($2 \tan \theta_{23}$). 
We show how the disappearance channel at the Neutrino Factory is complementary to the ``golden'' and 
``silver'' appearance channels and can be used to reduce the eightfold-ambiguity problem in ($\theta_{13}-\delta$).
\vspace{1pc}
\end{abstract}

% typeset front matter (including abstract)
\maketitle

%\section{The $\nu_\mu$ disappearance channel at T2K-I and the SPL}

Most of recent experimental breakthroughs in neutrino physics have been achieved
by exploiting the so-called ``disappearance channels'', i.e. counting how many $\nu_e$'s and $\nu_\mu$'s
are observed in a given detector having a certain (precise or approximate) knowledge of the neutrino fluxes
emitted from a given source. By observing a deficit in the neutrinos that reach the detector with respect
to those expected to be emitted from the source, a positive and eventually unambiguous signal of 
neutrino oscillations has been well established. 

New-generation experiments have been proposed to look for the unknown parameters $\theta_{13}$ and $\delta$ 
through ``appearance channels'', instead, such as $\nu_e \to \nu_\mu, \nu_\mu \to \nu_e$ (the ``golden channel'', \cite{Cervera:2000kp}) 
and $\nu_e \to \nu_\tau$ (the ``silver channel'',\cite{Donini:2002rm}). However, strong correlations between $\theta_{13}$ and 
$\delta$ and the presence of parametric degeneracies \cite{Barger:2001yr} in the ($\theta_{13},\delta$) parameter space make 
the simultaneous measurement of the two variables extremely difficult. 
A further problem arises from our present imprecise knowledge of atmospheric parameters, whose uncertainties
are far too large to be neglected when looking for such tiny signals as those expected in appearance 
experiments driven by the $\nu_\mu \to \nu_e$ and $\nu_e \to \nu_\mu, \nu_\tau$ oscillation probabilities \cite{Donini:2005rn}. 

\begin{figure}
\begin{tabular}{c}
\includegraphics[width=15pc]{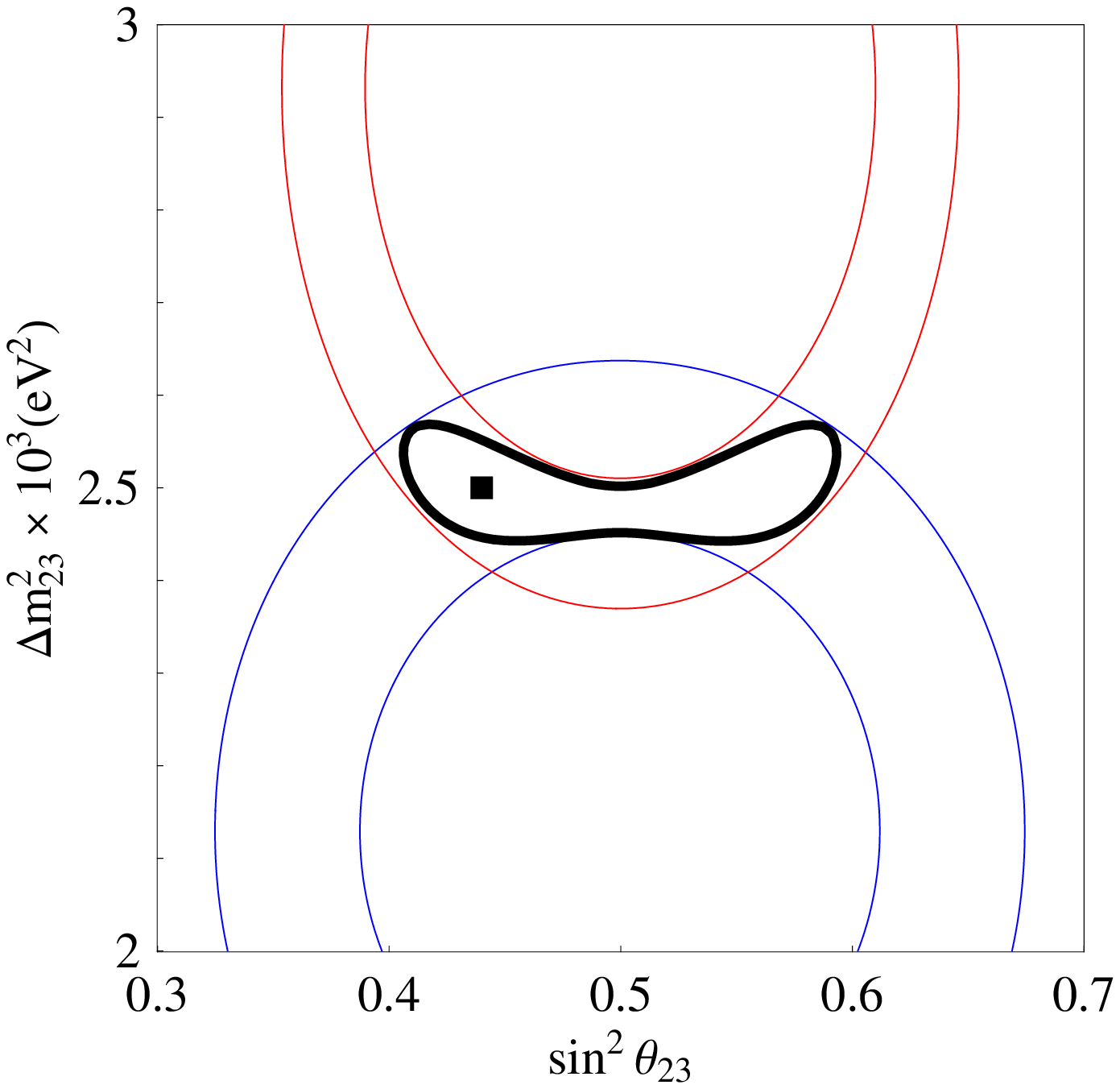} \\
\includegraphics[width=15pc]{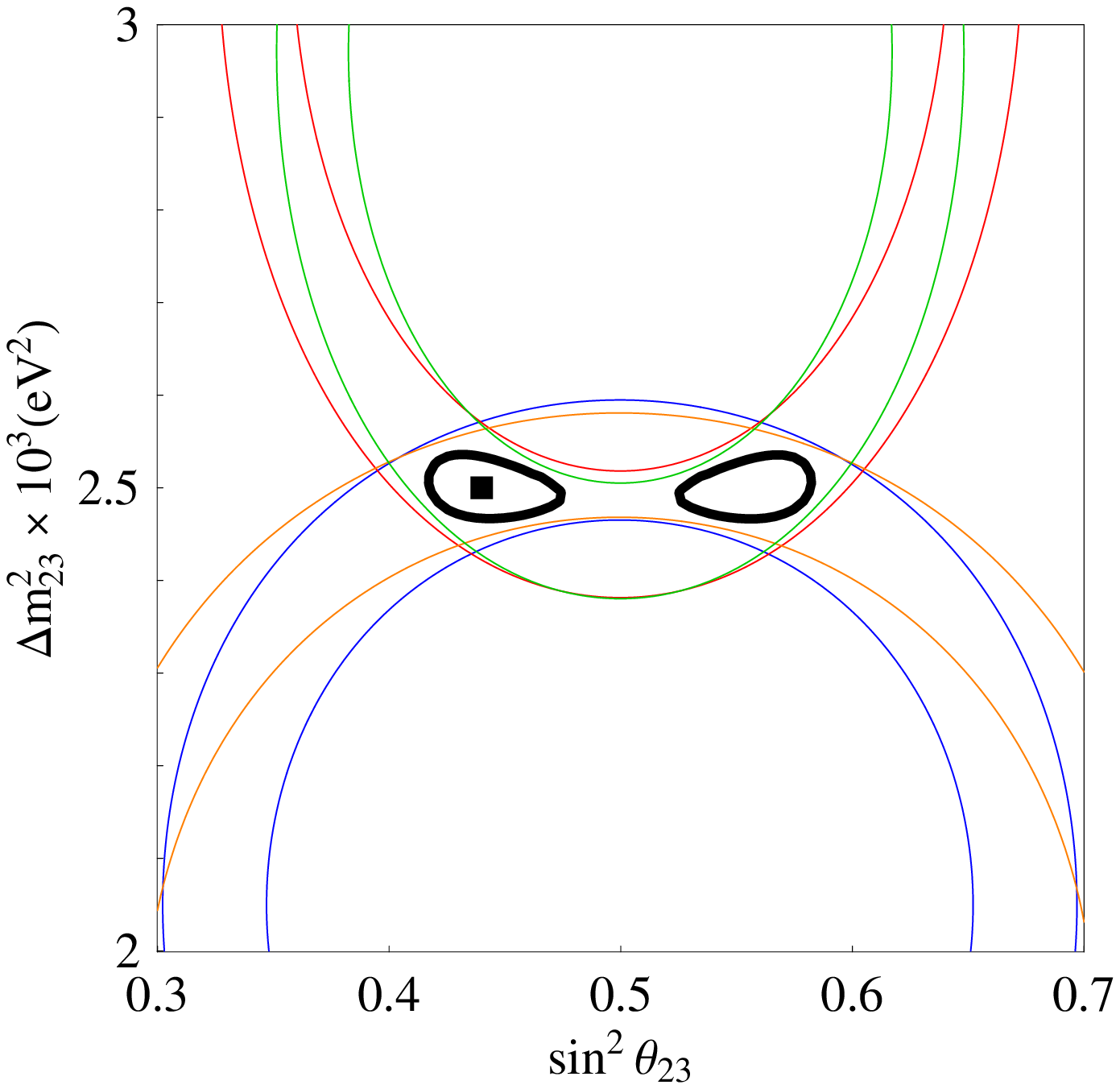} \\
\includegraphics[width=15pc]{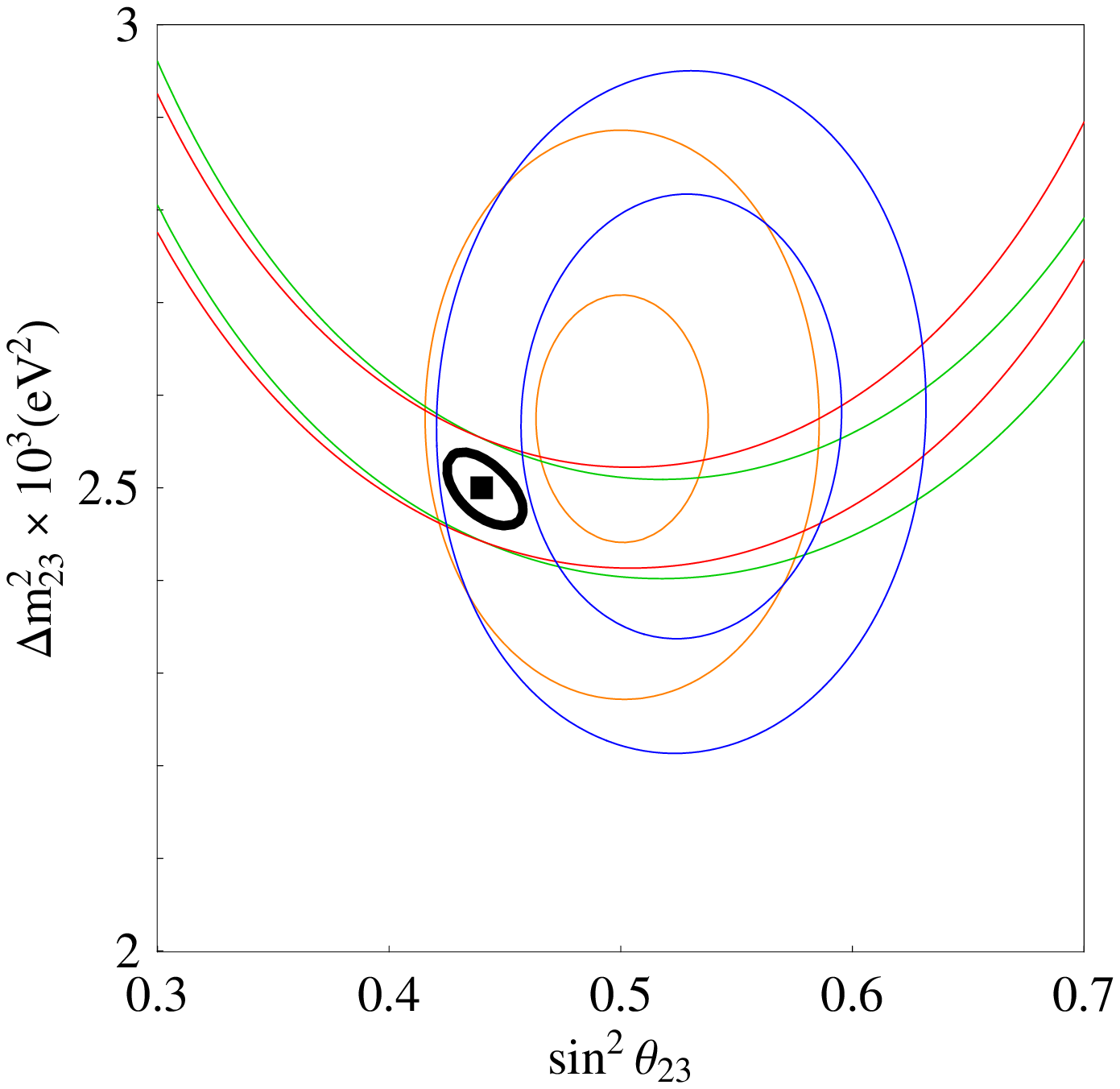}
\end{tabular}
\label{fig:bins}
\end{figure}

The $\nu_\mu$ disappearance channel is the best signal to reduce the uncertainties on atmospheric parameters \cite{Huber:2005ep}. 
In Fig.~\ref{fig:bins} we present a comparison of the 90 \% CL contours in the $(\theta_{23},\Delta m^2_{23})$
plane for $\nu_\mu \to \nu_\mu$ at T2K-I, the SPL and the Neutrino Factory with a $L = 3000$ km baseline.
Details on the different setups can be found in Ref.~\cite{setups}. 
The input parameters are: $\theta_{23} = 41.5^\circ$, $\Delta m^2_{23} = 2.5 \times 10^{-3}$ eV$^2$;
$\Delta m^2_{12} = 8.2 \times 10^{-5}$ eV$^2$, $\theta_{12} = 33^\circ$ and $\theta_{13} = \delta = 0^\circ$
(apart from Fig.~\ref{fig:bins}(bottom), with $\theta_{13} = 8^\circ$).

{\bf T2K-I}, Fig.~\ref{fig:bins}(top)\\
5 years of $\nu$'s divided into four 200 MeV bins with $E_{min} = 400$ MeV; only the first two bins are shown.
Notice that contours corresponding to neutrinos with an energy below and above the oscillation peak have a different shape. 
Thus, the combination of different bins significantly increases the $\theta_{23}$ resolution with respect 
to a counting experiment \cite{Donini:2004hu}. For non-maximal $\theta_{23}$ two degenerate solutions are found, 
$ \sim \theta_{23}$ and $\sim \pi/2 - \theta_{23}$. Taking into account our present ignorance on sign($\Delta m^2_{23}$), 
two more solutions are found at a different value of $\Delta m^2_{23}$ (that depends on the precise values of $\theta_{13}$ 
and $\delta$ \cite{Donini:2004hu}).
T2K-I can measure $\Delta m^2_{23}$ with a precision $\sim 10^{-4}$ eV$^2$ for $\Delta m^2_{23} = 2.5 \times 10^{-3}$ eV$^2$, 
but cannot exclude maximal mixing at 90 \% CL for these input parameters.

{\bf SPL}, Fig.~\ref{fig:bins}(middle) \\
2 years of $\nu$'s and 8 years of $\bar \nu$'s; two bins: $E_1 \in [0,250]$ MeV, $E_2 \in [250,600]$ MeV, \cite{Blondel:2004cx}.
The SPL can measure $\Delta m^2_{23}$ with a precision $\sim 10^{-4}$ eV$^2$ for
$\Delta m^2_{23} = 2.5 \times 10^{-3}$ eV$^2$ and can exclude maximal mixing at 90 \% CL for these input parameters.
Notice that a rather ``small'' experiment such as T2K-I, with 5 years of runtime with $\nu$ and a 22.5 kton detector, 
has a ($\theta_{23},\Delta m^2_{23}$)-resolution not much worse than the SPL, with 2+8 years of runtime with $\nu, \bar \nu$
and a 440 kton detector. One of the reasons is that the SPL has both $\nu$ and $\bar \nu$ beams with the same average energy 
(tuned to the first oscillation peak). Thus, information from the two beams add statistically 
but not complementarily (in the absence of matter effects), as it can be seen in the Figure.

{\bf NF@3000}, Fig.~\ref{fig:bins}(bottom) \\
5 years of $\nu$'s and $\bar \nu$'s divided in 4 GeV bins; only the on-peak and above-peak bins are shown.
Notice that the resolution in $\theta_{23}$ is extremely good and that maximal mixing can be easily 
excluded for $\theta_{23} = 41.5^\circ$. Moreover, when a non-vanishing $\theta_{13}$ is switched on matter effects become important 
and introduce a strong $\theta_{23}$-asymmetry that can be clearly seen in the Figure for $\theta_{13}=8^\circ$
(other parameters are the same as before). This asymmetry solves the disappearance octant ambiguity for $\theta_{13} \geq 3^\circ$.
Solving the octant-degeneracy in the disappearance channel (for large $\theta_{13}$) can be used to reduce the eightfold-degeneracy
of the appearance channels (see Ref.~\cite{Donini:2003kr}, also). 

%As a consequence, for generic values of the input pair ($\bar \theta_{13},\bar \delta$), we get eight different solution in 
%the ($\theta_{13},\delta$) plane. Different ways to solve the eightfold ambiguity have been proposed in the literature, 
%such as to combine experiments with different $L/E$ \cite{Burguet-Castell:2001ez}, or the golden channel at the Neutrino
%Factory with $\nu_\mu \to \nu_e$ at a SuperBeam \cite{Burguet-Castell:2002qx}, or the golden and silver channels 
%at the Neutrino Factory \cite{Donini:2002rm,Autiero:2003fu}, or even to combine different SuperBeams \cite{Minakata:2003ca} 
%or a SuperBeam and a $\beta$-Beam \cite{Bouchez:2003fy,Donini:2004hu}.
%In Ref.~\cite{Donini:2003kr} it was shown that the combination of the Neutrino Factory 
%with a SPL-like SuperBeam was able to solve the eightfold-degeneracy and to give a single allowed region 
%in the ($\theta_{13},\delta$) plane at the price of three specialized detectors, a 40 kton Magnetized Iron Calorimeter 
%and a 4 kton Emulsion Cloud Chamber (to look for $\nu_e \to \nu_\mu$ and $\nu_e \to \nu_\tau$ oscillations, respectively) 
%and a gigantic 1 Mton Water \v Cerenkov (to look for $\nu_\mu \to \nu_e$). 

\end{document}